\documentclass{article}

\usepackage{amsmath,amsfonts}
\usepackage{amsthm}

\usepackage{psfig}

\newcommand{\1}{\mathbb{I}}
\newcommand{\bra}[1]{\pmb{\langle}#1\pmb{|}}
\newcommand{\braket}[2]{
 \pmb{\langle}
 #1\pmb{\mid}#2\pmb{\rangle}}
\newcommand{\cfield}{\mathbb{C}}
\newcommand{\dff}{\scshape}
\newcommand{\eee}{\mathcal{E}}
\newcommand{\ee}{\mathbf{e}}
\newcommand{\h}{\mathbf{H}}
\newcommand{\ket}[1]{\pmb{|}#1\pmb{\rangle}}
\newcommand{\lth}{n}
\newcommand{\luders}[2]{{#1}_L\left[#2\right]}
\newcommand{\mynewdist}[2]{L\left(#1||#2\right)}
\newcommand{\rfield}{\mathbb{R}}
\newcommand{\rrh}{\rho}
\newcommand{\ssr}{\sigma}

\DeclareMathOperator{\const}{const}
\DeclareMathOperator{\support}{supp}
\DeclareMathOperator{\trc}{Tr}

\newtheorem{theorem}{Theorem}

\newcommand{\bracket}[3]{\bra{#1}{#2}\ket{#3}}
\newcommand{\ketbra}[2]{\ket{#1}\bra{#2}}
\newcommand{\raypr}[1]{\,\ketbra{#1}{#1}\,}
\newcommand{\scf}{\cfield{}S}

\newtheorem{lemma}[theorem]{Lemma}

\title{A note on continuous ensemble expansions of quantum
states}

\author{
Rom\`an R. Zapatrin\\
\small\itshape
Friedmann Lab. for Theoretical Physics,
SPb EF University,\\
\small\itshape
Griboyedova 30--32,
191023, St.Petersburg, Russia;\\
\small\rm
e-mail: zapatrin@rusmuseum.ru
}

\date{}

\begin{document}

\maketitle

\begin{abstract}
Generalizing the notion of relative entropy, the difference
between {\em a priori} and {\em a posteriori} relative entropy for
quantum systems is drawn. The former, known as quantum relative
entropy, is associated with quantum states recognition. The
latter---{\em a posteriori} relative quantum entropy is introduced and 
shown to be related with state reconstruction due to the following
property: given a density operator $\rho$, ensembles of pure
states with Gibbs distribution with respect to the defined
distance are proved to represent the initial state $\rho$ up to an
amount of white noise (completely mixed state) which can be made
arbitrary small.
\end{abstract}

In classical probability the {\dff relative entropy} (or
Kullback-Leibler distance) $S(\rrh||\ssr)$ of a distribution
$\rrh=\{p_1,\ldots,p_{\lth}\}$ with respect to another
distribution $\ssr=\{q_1,\ldots,q_{\lth}\}$ is defined as

\begin{equation}\label{ekullleib}
 S(\rrh||\ssr) \;=\; \left\lbrace
 \begin{array}{ll}
  -\sum_{k}p_k\log\left(\frac{p_k}{q_k}\right)
  &
  \mbox{ if $\support\rrh\subseteq\support\ssr$}
\\
  +\infty & \mbox{ otherwise} \\
 \end{array}
 \right.
\end{equation}

\medskip

\noindent Usually, this notion is generalized in quantum
information theory (see, {\em e.g.} \cite{ohyapetz} for a
review) by analogy with von Neumann entropy, namely, the sum is
replaced by operator trace. For two density operators $\rrh$
and $\ssr$ the quantum relative entropy $S(\rrh||\ssr)$ reads:

\begin{equation}\label{edefqre}
S(\rrh||\ssr) \;=\; \left\lbrace
\begin{array}{ll}
   -\trc\left[\rrh\left(\log{}\rrh -\log{}\ssr\right)\right] &
   \mbox{ if $\support\rrh\subseteq\support\ssr$}
\\
\\
  +\infty & \mbox{ otherwise}
\end{array}
\right.
\end{equation}

\paragraph{Relative entropy and state recognition.} Although
relative entropy does not satisfy the triangle inequality and
is therefore not a `true' metric, it is a nonnegative convex
function of $\{p_k\}$ and equals zero only if the distributions
are equal, $\ssr=\rrh$. Its relevance to distinguishing
probability distributions is vindicated by the Sanov's theorem
\cite{coverthomas} which states that the probability
$P_{N}(\rrh|\ssr)$ that the state $\ssr$ passes the test
determining if the state is $\rrh$ tends to

\begin{equation}\label{esanovclass}
 P_{N}(\rrh|\ssr)
 \;\rightarrow\;
 \const\cdot
 e^{-NS(\rrh||\ssr)}
\end{equation}

\noindent as $N$ (the number of checked samples of $\ssr$) tends
to $\infty$. A quantum analog of Sanov theorem was also proved
\cite{hi-pe}. This shows that both classical and quantum
relative entropy is an adequate tool to {\em recognize} states.

\paragraph{Relative entropy from the operationalistic
perspective.} In this note I give a quantum {\em
operationalistic} analog of Kullback-Leibler distance
\eqref{ekullleib}. For that, note that each probability $q_k$
can be treated in two ways

\begin{itemize}
\item {\em a priori}, that is as the probability of the event
associated with the value $q_k$ of $\ssr$ to occur
\item {\em a posteriori}, that is, as the probability of the
event associated with the value $p_k$ of $\rrh$ to occur
provided the system was in state $\ssr$ \end{itemize}

\noindent Coming to quantum formula according to {\em a priori}
interpretation, we get the standard quantum relative entropy
\eqref{edefqre} of $\rrh$ with respect to $\ssr$. Following the
second analogy, we get the relative entropy of $\rrh$ with
respect to the post-measurement state, which is called {\dff
L\"uders state} \cite{lueders}, it reads

\begin{equation}\label{edeflueders}
 \luders{\rrh}{\ssr}
 \;=\;
 \sum_{k=1}^{\lth}\limits
 \raypr{\ee_k} \ssr \raypr{\ee_k}
\end{equation}

\noindent and describes the state of the system which was
initially in state $\ssr$ after the measurement associated with
the density operator $\rrh$ was carried out. Now we can
introduce the distance between the states---the {\em a
posteriori} quantum generalization of relative entropy---as
follows

\begin{equation}\label{efirstdefmynewdist}
\mynewdist{\rrh}{\phi} \;=\; S\left( \rrh\rVert
\luders{\rrh}{\ssr} \right)
\end{equation}

\noindent Note that the value of this distance may be finite
when $\ssr=\raypr{\phi}$ is a pure state and $\rrh$ is
mixed, which is never the case for quantum relative entropy.
For further purposes the explicit expression---an analog of
\eqref{edefqre}---for the distance function
\eqref{efirstdefmynewdist} between a pure state $\phi$ and
mixed state $\rrh=\sum_{k=1}^{\lth}\limits p_k\, \raypr{\ee_k}$
is to be written down:

\begin{equation}\label{edefmynewdist}
   \mynewdist{\rrh}{\phi} \;=\; \left\lbrace
\begin{array}{ll}
 -\sum_{k}p_k\log\frac{p_k}{\lvert\braket{\ee_k}{\phi}\rvert^2}
 &
 \mbox{ if $\phi$ has nonzero overlap with $\rrh$}
 \\
 \\
 +\infty
 &
 \mbox{ otherwise}
\end{array}
\right.
\end{equation}

\noindent---a vector $\phi$ is said to have \cite{gdansk} a
{\dff nonzero overlap} with $\rrh$ if
$\bracket{\phi}{\rrh}{\phi}\neq 0$.

\medskip

\paragraph{Reconstructing quantum states by Gibbs ensembles.}
Given a full-range density operator $\rrh$ in
$\h=\cfield^\lth$, form a kind of Gibbs ensemble of pure states
treating $\mynewdist{\rrh}{\phi}$ as distance:

\begin{equation}\label{egibbsens}
    \mu_\rrh(\phi)
    \;=\;
    K \exp\left(-\frac{\lth}{\varepsilon}
    \mynewdist{\rrh}{\phi}\right)
\end{equation}

\noindent where $\varepsilon>0$ is a small parameter and $K$ is a
normalization constant with respect to Haar measure $dS$ on the
set $\scf^\lth$ of unit vectors in $\h$. This ensemble is
associated with a density operator

\begin{equation}\label{ebigmix}
    I(\rrh)
    \;=\;
    \int_{\scf^\lth}\limits \mu_\rrh(\phi) \, \raypr{\phi}
dS
\end{equation}

\noindent and the following {\dff reconstruction formula} holds:

\begin{equation}\label{emainformula}
 \int_{\scf^\lth}\limits \mu_\rrh(\phi)
 \raypr{\phi} dS
 \;=\;
 (1-\varepsilon)\rrh+\varepsilon\Lambda
\end{equation}

\noindent where $\Lambda=\frac{1}{\lth}$ is the density matrix of
white noise---a completely mixed state. The proof of this
formula is a routine integration, see Appendix for details and
the explicit expression for the normalization constant.

\paragraph{Limit distribution.} The limit distribution in
\eqref{emainformula} corresponds to an ensemble $\eee$ whose
density matrix is exactly $\rrh$, let us describe it
explicitly.  According to \eqref{emainformula}, the support of
the limit distribution is the set of unit vectors for which the
{\em a posteriori} distance from $\rrh$ is zero, that is,
\( \support(\eee)
=
 \{
 \phi_{\theta_1,\ldots,\theta_\lth}\mid
 \mynewdist{\rrh}{\phi_{\theta_1,\ldots,\theta_\lth}}=0
 \}
\) with

\[
 \phi_{\theta_1,\ldots,\theta_\lth}
 \;=\;
 \sum_{k=1}^{\lth}\limits
 e^{-\theta_k}\sqrt{p_k}\;\ket{\ee_k}
\]

\noindent where $\theta_1,\ldots,\theta_\lth=0\ldots 2\pi$.
Therefore the resulting distribution is uniform over the
$\lth$-torus $S^1\times\cdots\times S^\lth$ and, as a
consequence, any full-range density operator $\rrh$ can be
represented as the following {\dff uniform ensemble}

\begin{equation}\label{easympdistr}
 \rrh
 \;=\;
 \frac{1}{(2\pi)^\lth}
 \int_{0}^{2\pi}\limits
 \cdots
 \int_{0}^{2\pi}\limits
 \raypr{\phi_{\theta_1,\ldots,\theta_\lth}}
 d\theta_1\cdots d\theta_\lth
\end{equation}

\paragraph{An example.} Consider a density operator $\rrh$ whose
spectral decomposition is
\begin{equation}\label{enaitis}
\rrh=\frac{1}{4}\raypr{0}+\frac{3}{4}\raypr{1}
\end{equation}
\noindent in two-dimensional real space $\h=\rfield^2$. The
picture below shows three expansions of $\rrh$ (note that in case
of real state space the uniform ensemble \eqref{easympdistr} is
discrete as the phase multiples are $\pm 1$)

\bigskip
\medskip

\unitlength=1pt
\begin{center}
\begin{tabular}{ccc}
\psfig{figure=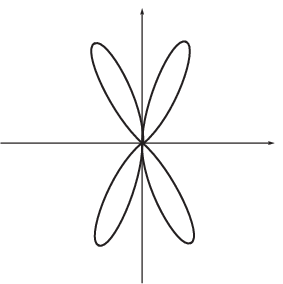} \qquad & \qquad \psfig{figure=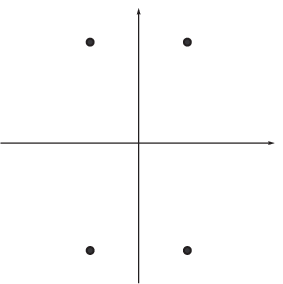}
\qquad & \qquad \psfig{figure=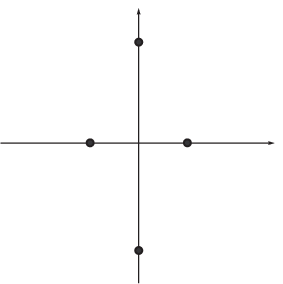} \cr Gibbs ensemble
\eqref{emainformula} & Uniform ensemble \eqref{easympdistr} &
Classical ensemble \eqref{enaitis}
\end{tabular}
\end{center}

\paragraph{Acknowledgments.} The attention to work offered by
the participants of the Friedmann seminar on theoretical
physics (St.Petersburg), in particular, to S.Krasnikov and
A.Lobashev is highly appreciated.

\newpage

\subsection*{Appendix: the proof of formula
\eqref{emainformula}}

Introduce the following numerical integral

\begin{equation}\label{edefippp}
I^{a_{1}\ldots{}a_{\lth}}_{\lth}
\;=\;
\int_{z\in\scf_\lth}\limits
\prod_{k=1}^\lth
|\braket{\ee_k}{z}|^{2a_k}
\,dS
\end{equation}

\noindent for which the following lemma holds

\begin{lemma}\label{leippp}
For $a_1,\ldots,a_{\lth}\ge 0$ and
$\braket{\ee_k}{\ee_j}=\delta_{kj}$

\begin{equation}\label{eippp}
I^{a_{1}\ldots{}a_{\lth}}_{\lth} \;=\;
\frac{2\pi^{\lth}\,\prod_{k=1}^{\lth}{}\Gamma(a_k+1)}{\Gamma
\left({\lth}+\sum_{k=1}^{\lth}{}a_k\right)}
\end{equation}
\end{lemma}

\begin{proof}
Represent $\cfield^{\lth}$ as real space $\rfield^{2\lth}$ with
coordinates $(r_1,\phi_1;\ldots;r_{\lth},\phi_{\lth})$ so that
$\braket{\ee_k}{z}=z_k=r_ke^{i\phi}$. Write down the integral
\eqref{edefippp} as

\[
I^{a_{1}\ldots{}a_{\lth}}_{\lth} \;=\;
(2\pi)^{\lth}\int_{r_1^2+\cdots+r_{\lth}^2=1}
\prod_{k=1}^{\lth} r_k^{2a_k}
\,dS
\]

\noindent Getting rid of $r_1=\sqrt{1-r_2^2-\cdots-r_{\lth}^2}$ we
obtain

\[
I^{a_{1}\ldots{}a_{\lth}}_{\lth}
\;=\;
(2\pi)^{\lth}\cdot
\int_{r_2^2+\cdots+r_{\lth}^2\le{}1}
\left(1-\sum_{j=2}^{\lth}r_j^2\right)^{a_1}
\left(\prod_{k=2}^{\lth}{}r_k^{2a_k+1} \,dr_k\right)
\]

\noindent Now, integrating over $r_{\lth}$, we obtain

\[
I^{a_{1}\ldots{}a_{\lth}}_{\lth}
\;=\;
(2\pi)^{\lth}\cdot
\int_0^1{}dr_{\lth}\;r_{\lth}^{2a_{\lth}+1}
\int_{r_2^2+\cdots+r_{{\lth}-1}^2\le{}1-r_{\lth}^2}
\left(1-r_{\lth}^2-\sum_{j=2}^{\lth-1}r_j^2\right)^{a_1}
\left(\prod_{k=2}^{{\lth}-1}r_k^{2a_k+1} \,dr_k\right)
\]

\noindent which, under the substitution
$\xi_k=r_k/\sqrt{1-r_{\lth}^2}$ for $k=2,\ldots,{\lth}-1$ and
$r_{\lth}=\sin\alpha$ reads

\[
I^{a_{1}\ldots{}a_{\lth}}_{\lth}
\;=\;
(2\pi)^{\lth}\cdot
\int_0^{\pi/2}{}d\sin\alpha\;\sin^{2a_{\lth}+1}\alpha
\cos^{2a_1}\alpha
\left(\vphantom{A^A}
\cos\alpha\right)^{2
\left(\sum_{k=2}^{{\lth}-1}a_k\right)+2({\lth}-2)}
\times
\]
\[
\times
\int_{\xi_2^2+\cdots+\xi_{{\lth}-1}^2\le{}1}
\left(1-\sum_{j=2}^{\lth-1}\xi_j^2
\right)^{a_1}
\left(\prod_{k=2}^{{\lth}-1}\xi_k^{2a_k+1} \,d\xi_k\right)
\;=\;
\]
\[
\;=\;
2\pi
\int_0^{\pi/2}
\sin^{2a_{\lth}+1}
\left(\vphantom{A^A}\cos\alpha\right)^{2
\left(\sum_{k=1}^{{\lth}-1}a_k+{\lth}-2\right)+1}d\alpha
\cdot
I_{{\lth}-1}^{a_{1}\ldots{}a_{{\lth}-1}}
\]

\noindent Using L\'egendre's formula

\begin{equation}\label{elegtrig}
\int_{0}^{\pi/2}
\sin^{2\alpha+1}x\,
\cos^{2\beta+1}x\,
dx
\;=\;
\frac{1}{2}\cdot
\frac{\Gamma(\alpha+1)
\Gamma(\beta+1)}{\Gamma(\alpha+
\beta+2)}
\end{equation}

\noindent we obtain the following recurrent expression

\begin{equation}\label{eippprecurr}
I^{a_{1}\ldots{}a_{\lth}}_{\lth}
\;=\;
\pi\cdot
\frac{\Gamma(a_{\lth}+1)\Gamma\left(
{\lth}-1+\sum_{k=1}^{{\lth}-1}a_k\right)}{\Gamma\left(
{\lth}+\sum_{k=1}^{\lth}a_k\right)}\cdot
I_{{\lth}-1}^{a_{1}\ldots{}a_{{\lth}-1}}
\end{equation}

\noindent Direct calculations show that the formula
\eqref{eippp} satisfies the recurrent
expression \eqref{eippprecurr},
and it remains to prove the induction
base. Do it for $\lth=1$:

\[
I_1^{p}
\;=\;
\int_0^{2\pi}d\phi\cdot{}1^p
\;=\;
2\pi
\]

\noindent which also accords with \eqref{eippp}. This completes
the proof.

\end{proof}

\bigskip

Now let $A=\sum_{k=1}^{\lth}a_k\, \raypr{\ee_k}$ be a
positive self-adjoint operator in $\h=\cfield^\lth$. Introduce
the operator integral

\begin{equation}\label{edefiaa}
I(A)
\;=\;
\int_{\scf^\lth}\limits
\left(
\prod_{k=1}^{\lth}
\lvert
\braket{\ee_k}{\phi}
\rvert^{2a_k}
\right)
\,
\raypr{\phi}
dS
\end{equation}

\noindent for which the following lemma holds.

\begin{lemma}\label{leaplusid}
For any positive self-adjoint operator $A$ in $\h$

\begin{equation}\label{e44}
I(A)
\;=\;
\frac{2\pi^{\lth}
\prod_{k=1}^{\lth}
\Gamma(1+a_k)}{\Gamma(1+{\lth}+\trc{A})}
\bigl(
A+\1
\bigr)
\end{equation}

\end{lemma}

\begin{proof}
First prove that $I(A)$ is diagonal in the eigenbasis of $A$.
For any $j\neq{}k$ the appropriate matrix element reads

\[
\bracket{\ee_j}{I(A)}{\ee_k}
\;=\;
(2\pi)^{\lth-2}
\iint_0^{2\pi}e^{i(\phi_k-\phi_j)}
d\phi_j
d\phi_k
\int_{r_1^2+\cdots+r_\lth^2=1}
\prod_i
r^{2a_i}
dS
\;=\;
0
\]

\noindent Now calculate the diagonal elements

\[
\bracket{\ee_k}{I(A)}{\ee_k}
\;=\;
\int_{\scf_\lth}
\left(
\prod_{i=1}^\lth
|\braket{\ee_i}{\phi}|^{2a_i}
\right)
\,
\left|\braket{\ee_k}{\phi}\right|^2
\,dS
\;=\;
I_\lth^{a_1\ldots(a_k+1)\ldots{}a_\lth}
\]

\noindent In the meantime it follows from \eqref{eippp} that

\[
I_\lth^{a_1\ldots(a_k+1)\ldots{}a_\lth}
\;=\;
I_\lth^{a_1\ldots{}a_\lth}
\frac{a_k+1}{\Gamma\left(
\lth+1+\sum_i{}a_i\right)}
\]

\noindent then $I(A)=\sum_k{}
\bracket{\ee_k}{I(A)}{\ee_k}
\cdot
\raypr{\ee_k}$, hence

\[
I(A)
\;=\;
\frac{2\pi^{\lth}
\prod\Gamma(a_i+1)}{\Gamma\left(
{\lth}+1+\sum{}a_i\right)}
\,
\sum_{k=1}^{\lth}
(a_k+1)
\raypr{\ee_k}
\]

\noindent which completes the proof.
\end{proof}

\bigskip

Denote by $\Lambda$ the density matrix of the white noise---a
completely mixed state

\[
\Lambda
\;=\;
\frac{1}{n}
\1
\;=\;
\frac{1}{n}
\sum_{k=1}^{\lth}
\raypr{\ee_k}
\]

\noindent then the exact version of formula
\eqref{emainformula} is stated by the following theorem.

\begin{theorem}\label{thexact}
Let $\rrh = \sum_{k=1}^{\lth}p_k\,\raypr{\ee_k}$ be a
positive ($p_k>0$) operator in $\h=\cfield^n$. Then for any
$\varepsilon\in(0,1)$

\begin{equation}\label{e44a}
(1-\varepsilon)\rrh+\varepsilon\Lambda
\;=\;
\frac{\Gamma\left(1+\frac{\lth}{\varepsilon}\right)
 e^{-\frac{\lth(1-\varepsilon)}{\varepsilon}
 S(\rrh)}}{2\pi^{\lth}
\prod_{k=1}^{\lth}
\Gamma\left(1+\frac{\lth(1-\varepsilon)p_k}{\varepsilon}
\right)}
\int_{\scf^\lth}\limits
e^{-\frac{\lth(1-\varepsilon)}{\varepsilon}
\mynewdist{\rrh}{\phi}
}
\,
\raypr{\phi}
dS
\end{equation}

\end{theorem}

\begin{proof} Consider the formula \eqref{e44} and let
$A=\frac{\lth(1-\varepsilon)}{\varepsilon}\rrh$, then
$a_k=\frac{\lth(1-\varepsilon)p_k}{\varepsilon}$ and

\[
\frac{2\pi^{\lth}
\prod_{k=1}^{\lth}
\Gamma(1+a_k)}{\Gamma(1+{\lth}+\trc{A})}
\left(
\frac{\lth(1-\varepsilon)}{\varepsilon}\rrh+\1
\right)
\;=\;
\int_{\scf^\lth}\limits
\left(
\prod_{k=1}^{\lth}
\lvert
\braket{\ee_k}{\phi}
\rvert^{2\frac{\lth(1-\varepsilon)p_k}{\varepsilon}}
\right)
\,
\raypr{\phi}
dS
\]

\noindent thus, taking into account that $\trc A =
\frac{\lth(1-\varepsilon)}{\varepsilon}$, we have

\[
\frac{\lth(1-\varepsilon)}{\varepsilon}\rrh+\1
\;=\;
\frac{\Gamma(1+
\frac{\lth}{\varepsilon})}{2\pi^{\lth}
\prod_{k=1}^{\lth}
\Gamma(\frac{\lth(1-\varepsilon)p_k}{\varepsilon}
+1)}
\int_{\scf^\lth}\limits
e^{\frac{\lth(1-\varepsilon)}{\varepsilon}
\sum_{k=1}^{\lth}
p_k
\log
\lvert
\braket{\ee_k}{\phi}
\rvert^{2}
}
\,
\raypr{\phi}
dS
\]

\noindent Using the definition \eqref{edefmynewdist} we obtain
\(
\frac{\lth(1-\varepsilon)}{\varepsilon}\rrh+\1\,=\)
\[\;=\;
\frac{\Gamma\left(1+\frac{\lth}{\varepsilon}\right)
 e^{\frac{\lth(1-\varepsilon)}{\varepsilon}
 \sum_{k=1}^{\lth}p_k\log{}p_k}}{2\pi^{\lth}
\prod_{k=1}^{\lth}
\Gamma\left(1+\frac{\lth(1-\varepsilon)p_k}{\varepsilon}
\right)}
\int_{\scf^\lth}\limits
e^{-\frac{\lth(1-\varepsilon)}{\varepsilon}
\sum_{k=1}^{\lth}
p_k
\left(
\log{}p_k
-
\log
\lvert
\braket{\ee_k}{\phi}
\rvert^{2}
\right)
}
\,
\raypr{\phi}
dS
\]

\noindent from which the formula \eqref{e44a} follows directly.
\end{proof}

\medskip

I emphasize that the formula \eqref{e44a} is exact. For small
$\varepsilon$ we assume $1-\varepsilon\simeq 1$ and obtain the
formula \eqref{emainformula}.


\begin{thebibliography}{99}

\bibitem{coverthomas} Cover, T. M. and Thomas, J. A.,
 {\itshape Elements of Information Theory.}
 New York: Wiley, 1991

\bibitem{hi-pe} F. Hiai, D. Petz,
 The proper formula for relative entropy and its asymptotics in
 quantum probability.
 {\itshape Communications in Mathematical Physics,}
 {\bfseries 143}
 99 (1991)

\bibitem{gdansk} M.Horodecki, Aditi Sen De, Ujjwal Sen,
 Quantification of quantum correlation of ensemble of states,
 quant-ph/0310100

\bibitem{lueders} G.L\"uders,
 \"Uber die Zustands\"anderung durch den Me\ss proze\ss,
 {\itshape Ann. Physik},
 {\bfseries 8}
 (6) 322 (1951);
 English translation `Regarding the state-change due to the
 measurement process' by K.A.Kirkpatrick is available at
 quant-ph/0403007

\bibitem{ohyapetz} M.Ohya, D.Petz,
 {\itshape Quantum entropy and its use},
 Springer, 1993

\end{thebibliography}
\end{document}